\documentclass[%
 aip,
 amsmath,amssymb,
 reprint,%
]{revtex4-1}

\usepackage{graphicx}
\usepackage{siunitx}
\usepackage{url}
\usepackage{dcolumn}
\usepackage{bm}
\usepackage{float}
\usepackage{textcomp,gensymb} 

\usepackage[utf8]{inputenc}
\usepackage[T1]{fontenc}
\usepackage{mathptmx}

\begin{document}

\preprint{AIP/123-QED}

\title[A low-energy MHz repetition rate short-pulse electron gun]{A low-energy MHz repetition rate short-pulse electron gun}

 \affiliation{Photon Science Institute, School of Physics and Astronomy, Faculty of Science and Engineering, University of Manchester, M13 9PL, UK.}

\author {Andrew James Murray}
\email{andrew.murray@manchester.ac.uk}

\author{Joshua Rogers}
\email{joshua.rogers@manchester.ac.uk}


\begin{abstract}
A pulsed electron gun that can produce MHz repetition rate nanosecond pulses is described. The gun uses a Pierce grid in combination with an anode to extract electrons from a tungsten filament cathode. The electrons emerging from the anode are accelerated and focused using two triple-aperture lenses to form a beam. By applying a high speed grid pulse that transitions through the extraction voltage region of the grid/anode combination, pulses of electrons are produced from the gun that have temporal widths less than 5 ns. The pulsed beams are produced at both rising and falling edges of the driving pulse. The characteristics of the emerging electron beams have been determined using an (e,2e) coincidence spectrometer and examples where they are used for time of flight decay measurements are given. 
\end{abstract}

\maketitle

\section{Introduction} \label{section:Introduction}

Pulsed excitation and ionization sources have wide application in areas ranging from electron spectroscopy through to mass spectrometry. Consider multi-photon excitation and ionization carried out using high-power pulsed lasers, where a sample is ionized prior to measuring the resulting fragments in a time of flight mass spectrometer \cite{ massspec_liquids, ZEKE, laser_TOF}. The laser system can either produce ionization directly through a non-resonant pathway, or a tuneable laser may be used to enhance ionization through different resonances in the target. In these experiments the laser pulses are short (ns, ps or fs) so that the time of ionization is well defined. Ions are then extracted and accelerated towards a detector, their time of flight yielding information about their mass.

By contrast, in electron spectroscopy the excitation mechanism is usually an electron beam because the ease with which the energy of the beam can be controlled makes them powerful excitation and ionization sources in many applications. A further advantage arises since electrons can remove angular momentum from the interaction when scattering at an angle to the incident beam, so that ${all}$ states of a target atom or molecule can then be excited. The normal selection rules that govern laser excitation hence do not apply. 

Electron beams are employed regularly in time-of-flight techniques where the momentum and kinetic energy of different products that may arise from an interaction with the beam are precisely controlled. Conventional coincidence experiments (where two or more fragments arising from an interaction with an electron beam are detected and then correlated in time) can be used to measure excited state lifetimes \cite{Imhof_Read_1977, WEDDING1990689}, to determine the angular `shape’ of an excited target \cite{ Andersen_Bartschat_2017} or to measure ionization differential cross sections in an (e,2e) reaction. These experiments mostly use a continuous electron beam, rather than a pulsed beam. 

Pulsing the incident electron beam produces a temporally well-defined onset to the interaction\cite{ BONHAM20071, Danica_Ca_Ions, aperture_pulsing_e2e1}, which unlocks additional information in electron spectroscopy experiments. Cold target recoil ion momentum spectroscopy (COLTRIMS) experiments adopt this technique by producing electron beams either directly \cite{COLTRIMS_gridpulsed} or by using ultra-fast lasers to produce an electron beam by photo-ionization \cite{ COLTRIMS_UVpulse, COLTRIMS_UVpulse_2}. Electron Momentum Spectroscopy (EMS) experiments may also adopt a pulsed source to probe the dynamics of the species under study \cite{electron_ion_coinc_Watanabe, pulsed_EMS_Takahashi, laser_EMS_apertured_gun}. The technique is exploited widely in velocity map imaging systems \cite{ VMI_DEA_1, VMI_SF6, DEA_KimballGun} as well as in other types of electron spectroscopy measurements \cite{TOF_H_H2, Morty_1, Morty_2, H2_dissociate, Absolute_CS}. 

Pulsed electron beams have also been used in target deflection experiments using supersonic beams \cite{Murray_Hammond_PRL, Murray_Hammond_RSI, Harries_2003}, in ion traps \cite{KLOSOWSKI201813} and for cold atom studies \cite{ AC_MOT, ColdAtom_Buckman}. Electron diffraction experiments have also used high energy pulsed electron beams to study the dynamics of reactions on surfaces \cite{ fslaserpulsegun, 30kVpulsedfsgun, Laserdriven}.

The principle mechanisms that are used to produce a pulsed electron beam in these experiments are either to use a laser source for the production of photo-electrons, or to modulate the beam current from an electron gun using electrostatic methods. Laser driven sources produce the shortest electron pulses, but these are expensive and require optical access for the laser beam to enter the vacuum apparatus. By contrast, electrostatically pulsed sources are compact systems that often require only modest modification of the driving electronics of a continuous electron gun to operate in this mode. It is this latter type of system that is described here.

Two main methods are used to electrostatically control the electron beam inside a gun. The simplest solution is to scan the electron beam through a small aperture inside the gun using a set of deflectors whose voltages are switched rapidly. When the beam passes through the aperture a pulsed beam is produced from the exit of the gun \cite{ Danica_Ca_Ions, aperture_pulsing_e2e1,Morty_3_apertured}. The switching time of the pulse then depends on the energy of the electron beam, the voltage applied to the deflector plates and their physical size. Since the electron beam is dumped inside the gun onto the back of the aperture plate, this method can produce scattered electrons that create a background flux of low-energy electrons and unwanted noise.

The second method uses a pulsed voltage applied to elements in the gun that are close to the cathode. Commercially available electron guns \cite{ Kimball_Guns} use this technique by applying a short pulse through a capacitor connected to a grid, or by directly driving the grid using a TTL driven power supply. These guns produce electron pulses with widths from 20~ns to 0.1~ms at a repetition rate of up to 5~kHz. They deliver pulses with rise and fall times of $\sim$10~ns. 

Zawadzki \textit{et. al.} \cite{ Morty_1, Morty_2} used a similar technique in the source region, however they included a small aperture ring between the cathode and anode whose voltage was switched to allow passage of electrons from the cathode to the anode. Their technique required a ns voltage pulse of around 40 Vpp to be applied to the ring through an isolating capacitor and impedance matching circuit. Their system operated at a rate of up to 500~kHz and produced pulses with widths between 2 – 3.5~ns. The Avtech pulse generator used in this system is however expensive ($\sim\$15,000$).

An alternative method of producing short pulses from an electron gun is presented here. This technique again exploits control of the electrons within the cathode region, by applying a control voltage to a Pierce grid located between the cathode and anode. Electron pulses with widths less than 5~ns can be produced from the gun at repetition rates of several MHz, using inexpensive electronics that can be built in the laboratory. Results from experiments using this gun are presented here to demonstrate its capabilities.

This paper is divided into 6 sections. Following this introduction the electron gun which is used in these experiments is described. The techniques discussed here can be used in any electron gun that uses a thermally heated cathode, grid and anode extraction. Section \ref{section:Grid Control} describes the electronic systems used to control the electron beam. Section \ref{section:width_measurements} details how the beam is controlled and gives examples of the pulses that are produced, as measured using a coincidence electron spectrometer. Examples of state decay measurements and how the gun operates for electron excitation studies using helium as a target are then presented in section \ref{section:experiments}, followed by conclusions and a summary of the technique in section \ref{section:Conclusions}.

\section{The Electron Gun} \label{section:Electron Gun}

\begin{figure} [ht]
    \centering
    \includegraphics[scale=0.55]{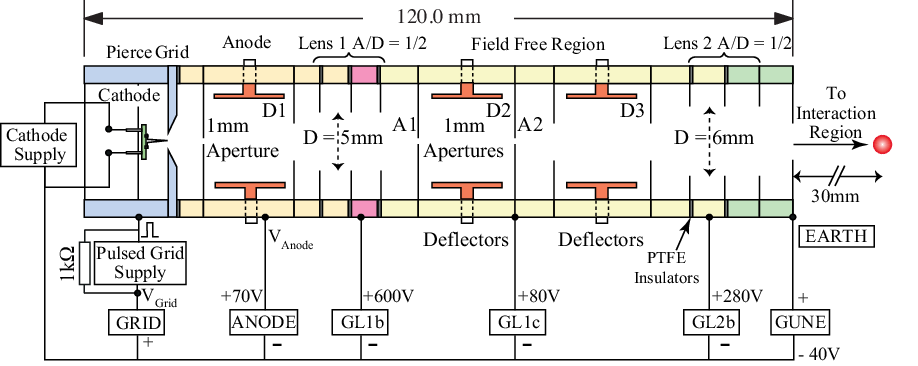}
    \caption{Schematic of the electron gun, which uses the combination of a Pierce grid and anode to extract electrons from the cathode region. Two aperture lenses (Lens 1 and Lens 2) in combination with defining apertures in the field free region then shape and accelerate the electrons to produce a beam at the interaction region that has zero beam angle and a pencil angle of $\sim$ 2$\degree$. The voltages shown are typical to produce a beam energy of 40~eV.}
    \label{fig:Electrongun}
\end{figure}{}

Figure \ref{fig:Electrongun} shows a schematic of the electron gun. Electrons are produced from a commercial tungsten filament (Agar A054) \cite{Agar} heated to emit thermal electrons by a cathode supply.  The cathode bias with respect to the interaction region is provided by the potential GUNE (Gun energy) and the supply sets the central tip of the filament at this potential. Electrons are extracted predominantly from the tip of the hot filament but some flux also emerges from adjacent parts of the filament. It is extremely difficult to model this region of the gun as space charge is high, localised magnetic fields may be present and the potential around the filament surface also varies strongly. 

The grid surrounding the cathode is of the Pierce design \cite{Pierce_Grid} and electron extraction from the cathode region is controlled by the combination of a negative potential $V_{\mathrm{grid}}$ on the Pierce grid \cite{Pierce_Grid} and a positive potential $V_{\mathrm{anode}}$ on the anode. The shape and potentials of the grid and anode then produce an electron beam that passes through the 1~mm anode aperture. 

The grid supply consists of a conventional DC supply `GRID' in series with the new pulsed grid supply that is described here. A 1~\si{\kilo\ohm} metal film resistor located at the grid inside the vacuum chamber serves as a load for this supply. 

Electrons passing into the anode region are imaged by lens~1 onto aperture A2 in the field free region set at the potential GL1c. Aperture A1 (positioned at the focal point of lens~1) then acts as the angle defining pupil for the electron optical system, constraining the pencil angle of the beam. Lens~2 images aperture A1 to infinity, so that the beam angle at the interaction region is zero. Aperture A1 is hence positioned at the focal point of lens~2. Each aperture has a diameter of 1~mm. By operating a two-stage electrostatic lens system as shown, the electron beam energy can be adjusted from $\sim$10~eV through to 300~eV, whilst maintaining zero beam angle at the interaction region located 30~mm beyond the exit of the gun. The diameter of the electron beam at the interaction region is typically around 1~mm over this energy range.

Deflectors positioned orthogonal to the beam axis are located in field free regions as shown, to correct for any displacement as the beam traverses through the gun. Deflectors D1 correct misalignment of the beam which passes through the anode aperture, and are referenced to the anode potential. Deflectors D2 between the defining apertures correct for effects due to space charge and surface charges on the lens elements in this region. Deflectors D3 steer the electrons onto the interaction region, thereby allowing for small mechanical misalignment in the gun. The potentials of deflectors D2 and D3 are referenced to the field free region voltage GL1c.

The continuous beam currents that can be delivered by the gun depend on the beam energy, as set by GUNE. At a beam energy of 40~eV, the gun can produce beam currents of up to 4~\si{\micro\ampere} at the interaction region when operating in continuous mode. This current is limited by space and surface charges inside the gun and along the beam. At an energy of 20~eV, the maximum beam current reduces to around 1~\si{\micro\ampere}, whereas at higher energies the gun can produce in excess of 10~\si{\micro\ampere}.

The electron gun is constructed from molybdenum, as this material has been found to produce the most stable electron guns for operation at low energy \cite{Comer_1971}. The gun consists of 20~mm diameter molybdenum tubing which has a wall thickness of 1~mm. The apertures are constructed from thin 80~\si{\micro\metre} molybdenum sheet. Insulating spacers between lens elements are made from PTFE rings of thickness 200~\si{\micro\metre}.

\begin{figure} [ht]
    \centering
    \includegraphics[scale = 0.65]{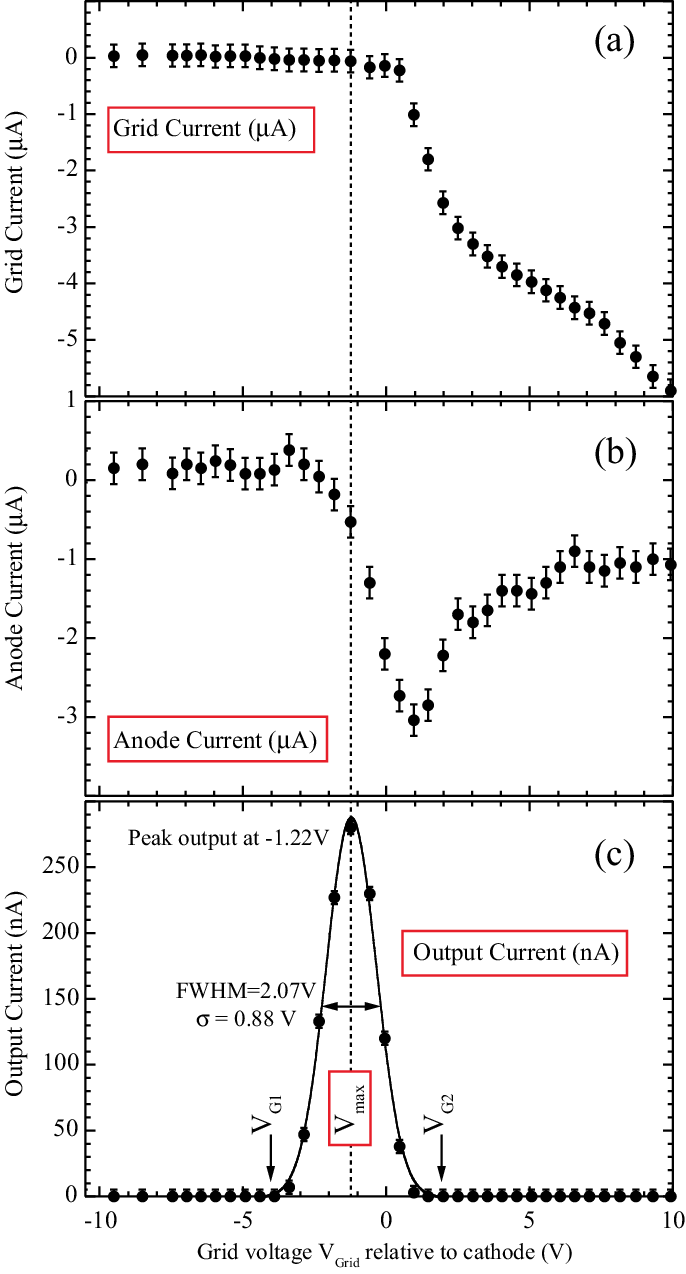}
    \caption{Measurement of emission currents impinging onto the grid and anode as well as the beam current measured on a Faraday cup beyond the interaction region, as a function of the grid voltage $V_{\mathrm{grid}}$. (a) shows the (negative) grid current, (b) shows the anode current and (c) shows the beam current reaching the Faraday cup beyond the interaction region.}
    \label{fig:Currentvsvoltage}
\end{figure}

To establish how the gun current varies as a function of the applied grid voltage, the output current was measured on a Faraday cup on the opposite side of the interaction region. The emission current from the cathode that reached the grid and anode was also measured. Figure \ref{fig:Currentvsvoltage} shows the result of this analysis where the grid voltage was changed from $V_{\mathrm{grid}}$ = -10~V to +10~V. Figure 2(a) shows that no current reaches the grid until it becomes positive with respect to the cathode, after which the (negative) grid current increases as electrons are attracted to it. The anode current remains around zero until the grid voltage reaches $V_{\mathrm{grid}}$ = $V_{\mathrm{G1}}$ $\sim$ -4~V, at which time electrons start to arrive at the anode. A beam current starts to emerge from the gun at this voltage. At a grid voltage $V_{\mathrm{grid}}$ = -1.22~V = $V_{\mathrm{max}}$ the output current reaches a maximum, beyond which it reduces back to zero at $V_{\mathrm{grid}}$ = $V_{\mathrm{G2}}$. The output current remains at zero as the grid voltage increases further. The anode current continues to increase until at $V_{\mathrm{grid}}$ = +1.0~V it reaches a maximum. The anode current then reduces, whereas the grid current continues to increase. At $V_{\mathrm{grid}}$ = +10~V $\sim$85\% of the emission current impinges on the grid, with only $\sim$15\% reaching the anode.

These results show that the beam current emitted from the gun is very sensitive to the grid voltage. In Figure \ref{fig:Currentvsvoltage}(c) a Gaussian is fitted to the data which has a full width at half maximum (FWHM) of around 2.07~V (standard deviation $\sigma$ = 0.88~V) centred around $V_{\mathrm{grid}}$ = $V_{\mathrm{max}}$. A SIMION model \cite{SIMION} of this region was unable to reproduce a peak in the output current, instead predicting a threshold grid voltage at which the beam current appears and above which it is largely unchanged. This failure of the model is likely because it does not include space charge. The response of the beam current to grid voltage seen in Figure \ref{fig:Currentvsvoltage}(c) is hence considered to be due to space charge effects within this region of the gun.

The data presented in Figure \ref{fig:Currentvsvoltage} shows how the gun can be operated in either a continuous or pulsed mode by modulating the grid voltage, while keeping unchanged all other voltages applied to the gun. For a continuous beam the grid is set to $V_{\mathrm{grid}}$ = $V_{\mathrm{max}}$, which produces maximum current from the gun. For long beam pulses that are on for a time $\Delta T_{\mathrm{long}}$, the grid can be set to a potential $V_{\mathrm{off}}$ below $V_{\mathrm{G1}}$ (beam off) and then rapidly switched up to $V_{\mathrm{max}}$, where it is held for the time $\Delta T_{\mathrm{long}}$, before it returns back to $V_{\mathrm{off}}$. 

For the shortest beam pulses from the gun the grid can be driven by a high slew rate voltage pulse that passes rapidly through the region between $V_{\mathrm{G1}}$ and $V_{\mathrm{G2}}$, in either a positive or negative direction. The slew rate of the driving pulse then dictates the temporal width of the emitted beam. It is this short-pulse mode of operation that is discussed here.

\section{Grid Control Electronics} \label{section:Grid Control}

To produce short-pulse beam currents from the electron gun as described above, it is necessary to produce controlled pulses from a high-speed amplifier that can slew rapidly through the region between $V_{\mathrm{G1}}$ and $V_{\mathrm{G2}}$. The amplifier must be biased so that its output $V_{\mathrm{off}}$ is below $V_{\mathrm{G1}}$ when no pulse is applied and is above $V_{\mathrm{G2}}$ when the pulse is switched on. At switch-off the amplifier output then slews back through $V_{\mathrm{G2}}$ and $V_{\mathrm{G1}}$ to settle at $V_{\mathrm{off}}$. Under these conditions the gun will then produce two short electron beams at twice the repetition rate of the driving pulse, with a time delay between these beams that is set by the driving pulse width.

High-speed operational amplifiers that satisfy these criteria can be obtained for less than $\$15$. Examples include the THS3491 current feedback amplifier \cite{THS3491} at $\sim\$14$ and the LM7171 voltage feedback amplifier  \cite{LM7171} at $\sim\$5$. The THS3491 has a maximum slew rate of 8~V/ns, whereas the LM7171 has a maximum slew rate of 4.1~V/ns. The minimum time to traverse between $V_{\mathrm{G1}}$ $\sim$ -4.1~V and $V_{\mathrm{G2}}$ $\sim$ +2V is hence $\sim$760~ps for the THS3491 amplifier and $\sim$1.5~ns for the LM7171. 

For the work discussed here the LM7171 was chosen as it requires lower power to operate. These very high speed amplifiers need carefully designed printed circuit boards (PCBs) to ensure stability when operating. They can deliver high output currents of up to 100~mA, which is beneficial when driving a coaxial feed to the grid in the gun.

\subsection{Pulse generation and TOSLINK optical feed} \label{subsection:pulser}

\begin{figure} [ht]
    \centering
    \includegraphics[scale=0.61]{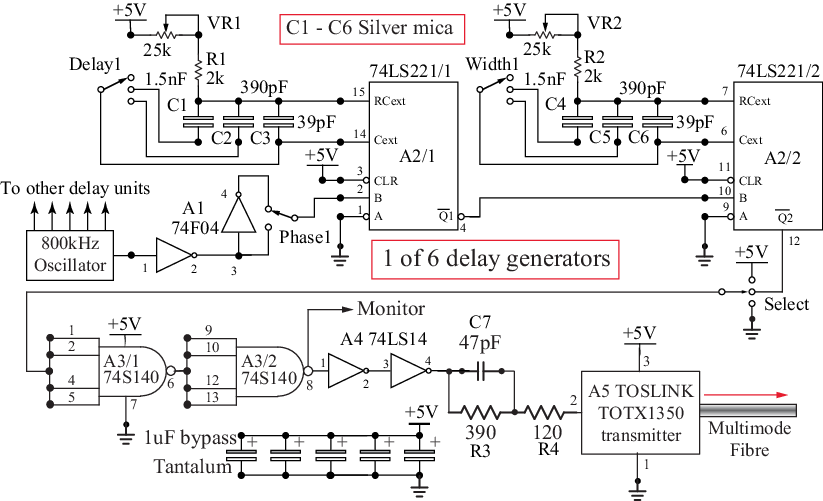}
    \caption{Schematic of the pulse driving circuit used in these experiments. An oscillator operating at 800~kHz switches the 74LS221 monostable A2/1 on, to produce a delayed output pulse at the output ${\bar Q_1}$, as set by the associated RC network VR1, R1, C1 – C3. The monostable A2/2 then produces an output at ${\bar Q_2}$ whose width is set by its associated network. The pulses drive a TOSLINK fibre transmitter through buffers A3 and A4. A multimode fibre links the signal to the grid drive electronics.}
    \label{fig:Grid_pulser}
\end{figure}

The design of this pulsed electron gun is part of a larger project that requires different components in an electron spectrometer and a laser to be switched sequentially at different times. To facilitate this a series of analog and digital delay generators have been designed based on previous work \cite{delay_generators}. Six analog delays have been built, together with five digitally controlled delays. Each delay generator is fed using an 800~kHz TTL clock generated from a 64~MHz master clock. The 800~kHz clock is buffered using a suite of 74S140 50~\si{\ohm} line drivers to match the impedance of RG58 cable that feeds the different delay units.

For the experiments discussed here one of the analog units was used to drive the grid pulse at 800~kHz, as shown in Figure \ref{fig:Grid_pulser}. This circuit consists of a delay generator followed by a variable width pulser. The delay time and pulse width are generated using cascaded 74LS221 monostable generators. The 800~kHz feed is buffered by 74F04 gates configured to switch either on the leading or falling edge of the clock signal via the `phase' switch. The RC network that sets the delay time consists of a 10-turn 25~\si{\kilo\ohm} potentiometer in series with a 2~\si{\kilo\ohm} metal film resistor that charges either capacitors C1, C2 or C3 as selected by a 3-position toggle switch. The output ${\bar Q_1}$ from this monostable then triggers the second monostable A2/2. The RC network for A2/2 is identical to that for the delay generator. A delayed variable-width 800~kHz TTL pulse is hence produced from the output at ${\bar Q_2}$. Silver mica capacitors are used for the timing control due to their excellent stability and very low temperature coefficient. 

The output from ${\bar Q_2}$ connects via a 3-position toggle switch (`select') to feed buffers A3/1 and A3/2. These are 74S140 50~\si{\ohm} line drivers, allowing the pulses to be monitored if required. The second line driver A3/2 feeds a pair of 74LS14 gates and a network consisting of C7, R3 and R4. This is the recommended drive for the TOTX1350 TOSLINK transmitter \cite{TOTX1350} which then feeds a multicore fibre. The TOSLINK interface has been chosen because it is used extensively in digital audio applications and so the fibre and couplers are inexpensive and easily obtained. The TOTX1350 transmitter is used because it is stable under both pulsed and DC operating conditions, so that the output can be selected to either be pulsed, off or on continuously, as set by the `select' switch. 

The oscillator and delay generators are all fed by a 5~$V_{\mathrm{DC}}$ supply that is reference to real earth. This allows easy interfacing to a personal computer or other peripherals, should this be required. The multimode fibre connection allows the amplifiers and switches driven by these circuits to be set to their local 0~V potential, which may be very different to real earth. As an example, for the grid control discussed here the local 0~V potential is set by the GUNE and GRID supplies as shown in Figure \ref{fig:Electrongun} and so may be up to 300~V below earth.    

\subsection{TOSLINK input and amplifier drive circuitry} \label{subsection:amplifier}

\begin{figure} [ht]
    \centering
    \includegraphics[scale=0.5]{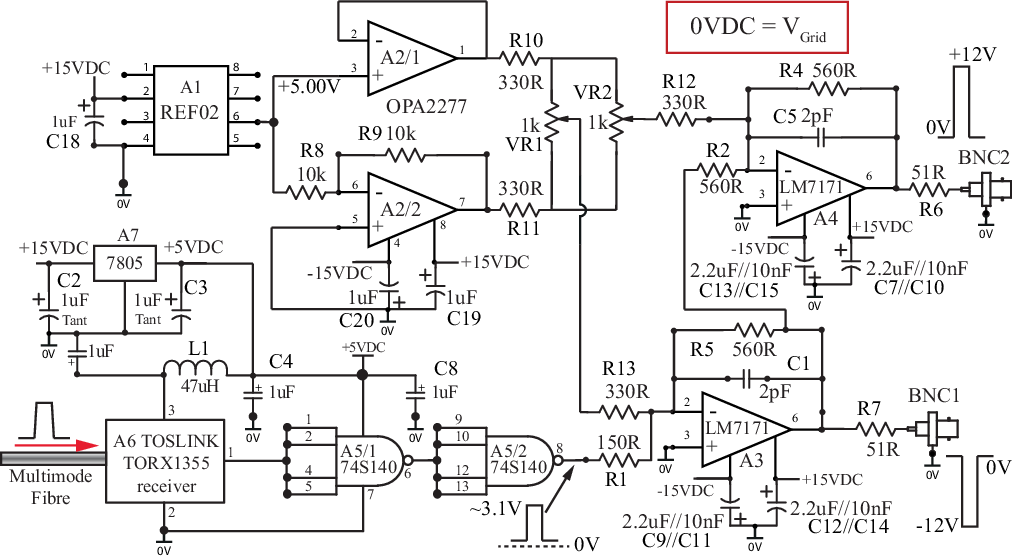}
    \caption{The grid voltage drive circuit. The TOSLINK receiver A6 is fed via fibre from the circuit shown in Fig. \ref{fig:Grid_pulser}. Output from the receiver is buffered by NAND gates A5 which feed amplifier A3. The output from A3 drives a second amplifier A4. A REF02 5~$V_{\mathrm{DC}}$ reference A1 buffered by amplifiers A2 provides an offset voltage to set the outputs to 0~V local using the trim potentiometers. The local 0~V for this circuit is referenced to $V_{\mathrm{grid}}$.}
    \label{fig:Grid_supply}
\end{figure}

Figure \ref{fig:Grid_supply} is a schematic of the grid drive circuit that supplies outputs from the BNC connectors. The optically encoded pulse is input via fibre to the TORX1355 receiver \cite{TOTX1350} A6 which feeds two 74S140 Schottky NAND gate line drivers. These gates produce fast TTL pulses that are in phase with the input signal. The output of gate A5/2 feeds amplifier A3, which is an LM7171 configured as an inverting amplifier and mixer with a gain of $\sim$3.7. The 2~pF feedback capacitor C2 in parallel with R11 reduces overshoot of the output pulse and stabilises the amplifier gain. The output of A3 connects to an external BNC1 via a 51~\si{\ohm} series resistor and also feeds a second inverting LM7171 amplifier A4 which has unity gain. A4 produces an amplified fast pulse via a second BNC2 that is in phase with the input signal.

The 74S140 gates produce fast TTL pulses which are slightly offset from 0V and which have a pulse height of $\sim$3.1~V. To compensate for the offset a REF02 5V reference is buffered by amplifiers A2/1 and A2/2 to produce $\pm$5~$V_{\mathrm{DC}}$ outputs that feed trim potentiometers VR1 and VR2. The outputs from the potentiometers then feed the summing inputs of the LM7171 amplifiers via R5 and R9. Any offset from A5/2 can hence be eliminated at both BNC outputs by setting the `select' switch in figure \ref{fig:Grid_pulser} to 0~V (ground) and then adjusting the trimmers until both outputs at BNC1 and BNC2 are zero. 

The pulsed outputs from BNC1 and BNC2 switch from 0~V local to around $\pm$12~V as shown. The rise and fall times of these outputs are set by the slew rates of both the 74S140 TTL gates and the LM7171 amplifiers. The pulsed signals are delivered to the grid through shielded coaxial cable from these BNC ports. It is however important to note that the shield of the cable is at the potential 0~V (local) set by the supply $V_{grid}$ - it is not at real earth and so cannot be connected directly to a normally grounded oscilloscope to check the pulses when the experiment is in operation. In the experiments described here the connection to the grid inside the vacuum chamber was through an isolated BNC connector on a conflat flange and the internal coaxial feed was connected directly to the grid. A 1~\si{\kilo\ohm} metal film resistor was connected across the feed at the grid, to return drive current back to the amplifier circuit as shown in Figure \ref{fig:Electrongun}.

The amplifier board is supplied from an isolated $\pm$15~$V_{\mathrm{DC}}$ regulated supply that also feeds the LM7805 +5~$V_{\mathrm{DC}}$ regulator A7 on the board. A7 supplies the TORX1355 receiver and TTL gates. The supplies to all active components are bypassed close to their supply pins using high frequency tantalum and ceramic capacitors. The TORX1355 has an additional 47~\si{\micro\henry} inductor in series with the supply, as recommended by the manufacturer.

For the experiments discussed in section \ref{section:width_measurements} the grid was fed by BNC2 so that the drive pulse was in phase with that from the pulse driving circuit. The pulses were set to have widths from $\sim$400~ns to 900~ns, depending on the measurement being made. The slew rate of the drive pulses was measured to be $\sim$1 - 2~V/ns using a fast oscilloscope. This is lower than the 4.1~V/ns of the LM7171 alone, due to the combination of gates and amplifiers as discussed above. 

To carry out measurements using the spectrometer the local 0~V reference was set to -6~$V_{\mathrm{DC}}$ with respect to the cathode using the GRID supply, as this was found to produce output pulses from the gun with the lowest width. No current was emitted when the pulse was set to 0~V, as expected from Figure \ref{fig:Currentvsvoltage}. Only a very small current was emitted when the pulse was high. As the drive pulse swept between $V_{\mathrm{grid}}$ = $V_{\mathrm{G1}}$ and $V_{\mathrm{G2}}$ in both rising and falling directions, a short burst of electrons was emitted.   

\section{Measurement of the emitted pulse widths} \label{section:width_measurements}

Measurement of the pulsed gun output characteristics was carried out using an existing (e,2e) spectrometer \cite{e2e_spectrometer}. This spectrometer has been used to carry out extensive surveys of electron-impact ionization of both atoms and molecules over a wide range of kinematic conditions \cite{e2e_1,e2e_2,e2e_3,e2e_4,e2e_5}. The spectrometer comprises an electron gun (as shown in Figure \ref{fig:Electrongun}), as well as two hemispherical electron analyzers, an atomic beam source and an internal photo-multiplier tube which monitors single photons emitted from the interaction region at $\sim$450~nm. The spectrometer is enclosed in a mu-metal lined vacuum chamber and all power supplies to the gun and analyzers are fully computer controlled and computer optimised. The experiment is capable of measuring scattered electrons with energies from 0.5~eV to 100~eV and the gun can produce electrons whose energy ranges from $\sim$5~eV to 300~eV.

Initial experiments on the gun characteristics were performed by introducing a helium beam into the interaction region with the electron beam emitting continuously (CW operation), so that the spectrometer could be optimised. The grid was hence set to a constant voltage $V_{\mathrm{grid}}$ = $V_{\mathrm{max}}$ and the pulser `select' switch was set to ground. The analyzers were adjusted to a residual energy of 18~eV at an angle of 60$\degree$ to the incident beam direction. An energy loss spectrum was then obtained by scanning the gun energy from 37~eV through to 42~eV (see Fig. \ref{fig:Elossspectrum}(a) below). 

To test and optimise the gun in pulsed mode the incident electron energy was set to the largest 2~$^1$P$_{1}$ peak of the inelastic spectrum at 39.2~eV and the GRID voltage adjusted to -6~$V_{\mathrm{DC}}$ with respect to the cathode. This reduced the scattered electron count rate into the analyzer to zero, as expected from Figure \ref{fig:Currentvsvoltage}. The grid pulsing circuit was then engaged, so that electron counts were once more detected from the interaction. A Time to Amplitude Converter (TAC) was triggered using a pulse from the 800~kHz oscillator that was converted from TTL to a negative-going NIM pulse. The TAC stop signal was obtained from the single electron counts that were being detected. The TAC output fed a Multichannel analyzer (MCA) \cite{e2e_6} which recorded the time of arrival of the electrons referenced to the 800~kHz trigger pulse.

\begin{figure}
    \centering
    \includegraphics[scale = 0.575]{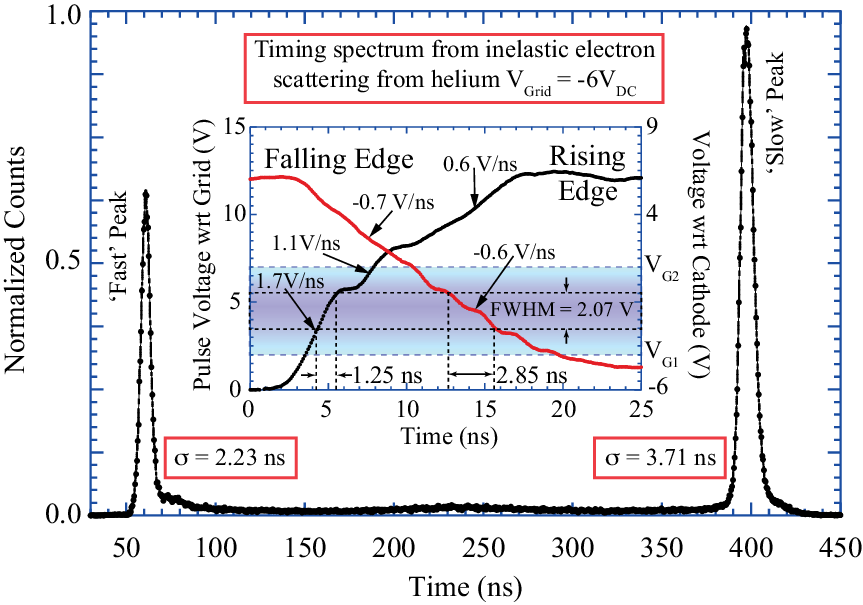}
    \caption{Accumulated timing spectrum where the electron beam was scanned over the inelastic scattering region of helium from 37~eV to 42~eV, the analyzer residual energy being 18~eV. The grid drive pulse was 350~ns wide and the TAC window was 500~ns. The delay was set so that both pulses of electrons appear within the timing window of the TAC as shown. The inset figure shows the rise and fall times of the grid driving pulse, measured on a fast (2~GHz) oscilloscope. The scale on the left is the grid voltage with respect to the cathode. The shaded region is between $V_{\mathrm{G1}}$ to $V_{\mathrm{G2}}$ where emission occurs (Fig. \ref{fig:Currentvsvoltage}(c)). The rising edge (which produces the `fast' peak) has a slope of 1.7~V/ns in this region. The emission peak with FWHM of 2.07~V hence predicts an equivalent timing peak with a FWHM of $\sim$1.25~ns ($\sigma_{\mathrm{fast}}$ $\sim$520~ps). The fall rate producing the `slow' peak has a negative slope of $\sim$0.6~V/ns between $V_{G1}$ to $V_{G2}$ and so the `slow' timing peak is predicted to have a FWHM of 2.85~ns ($\sigma_{\mathrm{slow}}$ = $\sim$1.2~ns).}
    \label{fig:Timingspectrum}
\end{figure}

Figure \ref{fig:Timingspectrum} shows a normalised timing spectrum which was accumulated over several hours, where the electron gun was slowly scanned from 37~eV through to 42~eV. The grid pulse was adjusted to be 350~ns wide and the TAC window was 500~ns. The pulse delay was set so that the signal from both the positive and negative transitions of the grid driving pulse appeared within the TAC timing window. The rising edge peak is labelled as the `fast' peak and the falling edge peak, the `slow' peak.

{}

The width of each peak was measured more precisely by placing a 50~ns TAC window over each, allowing a Gaussian to be fitted to them. The standard deviation, $\sigma$, of the fitted Gaussians is given in Figure \ref{fig:Timingspectrum}. The fast peak has a standard deviation of $\sigma$~=~2.23~ns and the slower peak has $\sigma$~=~3.71~ns.

The widths of the peaks in Figure \ref{fig:Timingspectrum} are not a direct measurement of the width of electron pulses from the gun but rather a convolution of the electron pulse duration with the temporal response of the electron analyzer. A SIMION model of the analyzers was hence carried out to estimate this contribution at the detected electron energy of 18~eV. This model predicts a variation of $\pm$360~ps in the flight time due to the input zoom lens. The pass energy of electrons entering the hemispheres was 15~eV for the data shown in Figure \ref{fig:Timingspectrum}. The region between the hemispheres was also modelled using more than 61,000 simulated trajectories over a range of energies across the input aperture, to determine the temporal distribution of electrons that exited the hemispheres. A bimodal distribution was found, with a total width of $\sim$ $\pm$2.3~ns. 

Considering both the zoom lens and hemispheres, the temporal spread of electrons inside the analyzers is hence estimated from the SIMION model to be $\sim$$\pm$2.4~ns. This is comparable to the measured width of the `fast' peak in Figure \ref{fig:Timingspectrum}, indicating that the actual width of the pulse emitted from the gun is considerably narrower than this and we refrain from deconvolving one from the other. The inset in Figure \ref{fig:Timingspectrum} shows that the predicted `fast' pulse standard deviation derived from the rise-time of the pulse should be $\sigma_{\mathrm{fast}}$ $\sim$520~ps, which is in agreement with these observations. By contrast, the grid pulse fall rate is predicted to produce a `slow' electron pulse from the gun with $\sigma_{\mathrm{slow}}$ $\sim$1.2~ns. Deconvolving the estimated temporal response of the analyzers from the measured width of the slow peak yields an value for the 'slow' electron pulse width of $\sigma$~$\sim$2.7~ns, which compares reasonably with the prediction from the inset of Figure \ref{fig:Timingspectrum}.

\section{Experimental data using the gun} \label{section:experiments}

To test the pulsed electron gun a number of different experiments were performed. These included measuring an energy loss spectrum from a gas target using a single analyzer, to making time of flight decay measurements of photons emitted from a target, where the incident electrons defined the start time for exciting the states. In both cases the (e,2e) spectrometer described above was used. 

\subsection{Energy loss measurements from helium} \label{subsection:Eloss}

\begin{figure} [!h]
    \centering
    \includegraphics[scale=0.720]{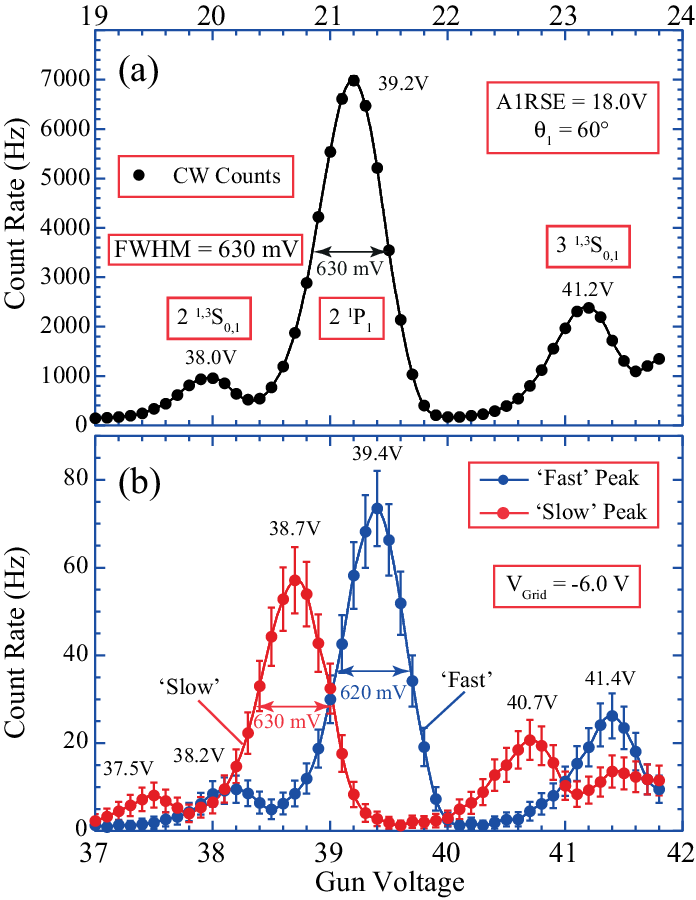}
    \caption{Energy loss spectrum from helium with the electron detector set to a residual energy of 18~eV and a scattering angle of 60$\degree$. (a) The spectrum taken with the electron gun operating in continuous (CW) mode. The states of helium excited by electron impact are shown, the peaks having a width of $\sim$630~meV. The scale above the figure shows the excitation energy of the states. (b) Results from the individual `fast' and `slow' peaks as the electron gun  was scanned over the same energy range. The energy-loss spectra occur at different gun voltages as shown, indicating that the energies of the electrons for each pulsed emission are different. The difference in their relative energies is $\sim$700~meV and they lie either side of the energy of the electrons emitted in CW mode.}
    \label{fig:Elossspectrum}
\end{figure}{}

Figure \ref{fig:Elossspectrum} shows an energy loss spectrum taken with the spectrometer operating both in the conventional continuous beam (CW) mode and when the gun was pulsed. Helium was chosen as the target gas and the electron analyzer was set to detect scattered electrons with 18~eV energy from the interaction, at a scattering angle of 60$\degree$. The electron gun was set to run in CW mode (Figure \ref{fig:Elossspectrum}(a)) with an output current of 120~nA. The operating pressure in the spectrometer was held at $\sim~7$~x~10$^{-6}$~Torr during these measurements, the baseline pressure in the chamber being $\sim~6$~x~10$^{-7}$~Torr.

The electron gun voltage GUNE was scanned from 37~V to $\sim$42~V as shown, with the grid set to $V_{\mathrm{grid}}$ = -1.2~$V_{\mathrm{DC}}$. The peaks in the spectrum occur due to the incident electron losing energy when it excites the atom. The lowest peak at 38.0~V is due to the (unresolved) 2~$^{1,3}$S states with excitation energies of 19.82~eV and 20.62~eV respectively. The 2~$^1$P$_{1}$ has an excitation energy of 21.2~eV and has the largest excitation probability under these conditions. The unresolved 3~$^{1,3}$S states have corresponding energies of 22.72~eV and 22.92~eV.

Figure \ref{fig:Elossspectrum}(b) is taken over the same energy range with the gun operating in pulsed mode. The grid bias was reduced to $V_{\mathrm{grid}}$ = -6~$V_{\mathrm{DC}}$, which set the 0~V reference for the pulse electronics. The data was collected by initially setting GUNE~=~37~V, then acquiring a full temporal profile (like that shown in Figure \ref{fig:Timingspectrum}) for 500~s. The gun voltage was then increased by 0.1~V and the process repeated until a full spectrum was accumulated. The individual temporal profiles were analysed to determine the area under both the `fast' and `slow' peaks at each energy. The resulting data were then normalised to produce the individual spectra in Figure \ref{fig:Elossspectrum}(b).

The data in Figure \ref{fig:Elossspectrum}(b) show that the inelastic spectra occur at different gun voltages GUNE for the `fast' and `slow' peaks. Since the detected electron energy was fixed at 18~eV, this demonstrates that the emitted electrons within each timing peak have different energies and are separated in energy by ~$\sim$0.7~eV. They also have a different energy to that of the CW beam, electrons emitted in the `fast' peak being shifted by +0.2~eV from those emitted in the CW beam. The energy loss peaks in the pulsed spectra have a similar width to those in the CW spectrum, indicating that the measured energy offset affects the pulsed emission as a whole.

This result is interesting as the energy of the electron beam emitted from the gun is set mostly by the potential difference between GUNE and the interaction region, which is at earth potential. A distribution of energies around the mean is expected due to thermionic emission from the cathode and due to space charge effects in the cathode region. Pulsing the grid as described here will not affect thermionic emission in any significant way. The differences between CW and pulsed modes seen in Figure \ref{fig:Elossspectrum}  hence must be due to changes to the space charge in this region. 

Further analysis shows that these energy differences are independent of the overall width of the grid pulse that sets the time delay between `fast' and `slow' pulses. They hence appear to be driven by dynamic changes to the space charge as the driving grid pulse transitions from 0~V to its peak value and back again. As noted above, space charge effects are extremely difficult to model physically and so the experimental observations are presented here without further analysis.    

\subsection{Time of flight decay measurements} \label{subsection:TOF}

The (e,2e) spectrometer is not setup to allow direct time of flight measurements for ions produced in the interaction region, however a fast bialkali photo-multiplier tube (PMT) is installed inside the spectrometer to allow photons from the electron-target interaction to be detected. Signals from the PMT are normally used to steer and focus the electron beam onto the interaction region under computer control, independently of the electron detectors \cite{e2e_spectrometer}. The PMT is focused onto the interaction region using a 25~mm focal length lens and a 450~nm broadband interference filter is used to select blue radiation emitted from electron-excited target atoms. A broadband filter is adopted for this application as many atoms emit light in this region when excited.

Since the PMT was already installed in the spectrometer this was a convenient detector for testing the pulsed gun, by carrying out time of flight decay measurements for photons emitted by excited atoms. The gun was hence setup in pulsed mode and electrons from the `fast' pulse were selected as the excitation source. The grid pulse width was set to 900~ns so that the `fast' and `slow' pulses were separated by this time interval. The TAC start pulse was triggered from the 800~kHz clock as before. The TAC stop pulse was however now taken from the single photon count signals obtained from the PMT, rather than from the electron detectors. The TAC was set to 1000~ns and the delay generator adjusted so that the accumulating signal from the PMT was positioned within the MCA timing window. 

\begin{figure} [!h]
    \centering
    \includegraphics[scale = 0.7]{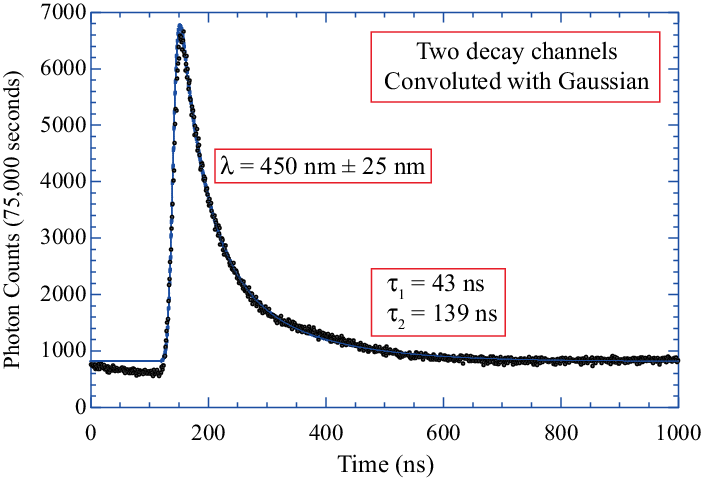}
    \caption{Measurement of the photon decay signal from electron impact excitation of helium.  The `fast' pulse was selected as the excitation source; the `slow' pulse was omitted from the measurement window but its tail can be seen in the baseline just before the `fast' pulse. The photons were selected using an interference filter with a peak transmission at 450~nm and a width of $\pm$25~nm. The data is fitted to two exponential decays with time constants of 43~ns and 139~ns, that are convolved with a Gaussian representing the temporal width of the incident electron beam and photon detector.}
    \label{fig:He_photon_decay}
\end{figure}{}

Figure \ref{fig:He_photon_decay} shows the result of these measurements, again using helium as the target gas. The incident electron energy was set to 30~eV and the gas jet was adjusted so that the chamber pressure was $\sim~2$~x~10$^{-6}$~Torr. The incident electron energy was chosen so that no He$^{+}$ states were excited. The TAC signal was accumulated for 21 hours, resulting in the decay curve shown. The signal was fitted to two exponential decay functions convolved with a Gaussian of width $\sim$7~ns, used to represent the temporal width of the PMT signals. A single exponential decay did not produce a good fit to the data. The decay times determined from the least squares fit to photon emission from the target were $\tau_1$ = 43~ns and $\tau_2$ = 139~ns.

The excited states in helium that produce radiation in the region from 400~nm to 500~nm are catalogued in the National Institute of Standards and Technology (NIST) database \cite{NIST_ASD}. This database shows that 13 transitions occur in this region, the strongest being the ${4^3}D\xrightarrow{{447.3\,\,{\mathrm{nm}}}}{2^3}P$ transition with a lifetime of 41~ns. This is the transition that produces radiation fitted to the data with a lifetime $\tau_1$. Five other transitions are relatively strong, with line strengths ranging from 10\% to 25\% of the 447.3~nm transition. These are the ${5^3}D\xrightarrow{{402.7\,\,{\mathrm{nm}}}}{2^3}P$ ($\tau$=82~ns), ${5^3}S\xrightarrow{{412.2\,\,{\mathrm{nm}}}}{2^3}P$ ($\tau$=224~ns) and ${4^3}S\xrightarrow{{471.4\,\,{\mathrm{nm}}}}{2^3}P$ ($\tau$=105~ns) triplet transitions, as well as the ${4^1}D\xrightarrow{{492.3\,\,{\mathrm{nm}}}}{2^1}P$ ($\tau$=50~ns) and ${5^1}D\xrightarrow{{438.9\,\,{\mathrm{nm}}}}{2^1}P$ ($\tau$=111~ns) singlet transitions. Their individual contributions are not possible to resolve with the filter used here but $\tau_2$ adequately describes their collective effect. 

The conventional method of measuring state lifetimes using a CW electron beam is to detect an electron that excites and scatters from the target in coincidence with an emitted photon \cite{Imhof_Read_1977,WEDDING1990689,Andersen_Bartschat_2017}. These electron-photon coincidence experiments are difficult and slow due to the relatively low count rates in the electron detector, which may be thousands of times smaller than the pulse rates adopted here. As an example, the typical electron detection rate when exciting a triplet state in helium is $\sim$1-2~kHz as shown in Figure \ref{fig:Elossspectrum}. Under these conditions it would take from 1 to 2 years to produces an equivalent signal to that presented in Figure \ref{fig:He_photon_decay}. This demonstrates one advantage of producing the short, high repetition rate pulses from the gun as described here. 

Although high repetition rate pulsed laser excitation of atoms can also be adopted for lifetime studies, these laser-based experiments are limited by selection rules governing excitation from the ground state. These limitations do not apply for excitation by electron impact, which allows all target states to be excited and studied.

\section{Conclusion} \label{section:Conclusions}

In this paper a new method has been described to produce high repetition rate pulses from an electron gun, with temporal widths $\sim$1~ns. The emission is controlled by applying a high slew rate voltage pulse to a Pierce grid surrounding the cathode. The drive electronics described here can be built for less than \$100, which makes the design affordable for most laboratories including those used for undergraduate teaching. Details of the design and board layout are available from the authors on request.

The pulsed gun has been tested over a wide range of beam energies from $\sim$5~eV to over 100~eV by detecting elastic and inelastic scattering from different target atoms in an electron spectrometer. These data are not presented here, however good pulse definition was found at all energies. Selected results from helium have been presented to demonstrate how the gun operates. These data show that the pulsed beams have slightly different energies to that of a CW beam and that the `fast' and `slow' pulse beams differ in energy by $\sim$0.7~eV. The reasons for this difference are not fully understood, but are thought to be due to changes in the space charge region between the cathode and anode as the grid pulse changes state. The energy difference between `fast' and `slow' beams does not depend on the time delay between them and so this could be exploited in pump-probe studies where the time difference is varied to make a measurement.

The experiments described here were conducted using a pulse repetition rate of 800~kHz since this is required in new experiments being designed in Manchester. Higher rates of several MHz have also been tested and no degradation to the emitted pulses was observed. There is however an upper limit for the TOSLINK transmitters and receivers of 10~MHz. 

The electron gun can also be set to emit only `fast' or `slow' pulses by controlling one of the deflector sets in the gun (see Figure \ref{fig:Electrongun}), or by installing additional deflectors at the exit of the gun. The circuit described in Figure \ref{fig:Grid_pulser} is designed to produce complementary positive- and negative-going output pulses. The outputs from a second pulsing circuit can hence be connected to two opposing deflector plates, with the local 0V for the amplifiers set to the reference voltage of the deflector supply. A second delay generator driving this amplifier board can then be adjusted to selectively pass either the `fast' or `slow' pulses through to the output, while preventing emission of the secondary pulse in each cycle.

Pulsed electron beams have a wide range of uses in physics, chemistry, engineering and material science as demonstrated by the research papers referenced here. Applications range widely from time-of-flight measurements through to the injection of pulsed electron beams into accelerators. The simple and inexpensive method of producing short-pulse electron beams described here is hence expected to benefit a wide community of researchers.  

Full details on the hardware described in this paper can be obtained from the authors on request.

\section{Acknowledgements}

We wish to thank the Engineering and Physical Sciences Research Council (EPSRC) for funding this work through grants EP/W003864/1 and EP/V027689/1.  

\section{Authors' Contributions}
All authors contributed to the work presented in this paper. 

\section{Data Availability Statement}
The data, schematics and PCB layouts that support the findings of this study are available from the corresponding author upon reasonable request.

\bibliography{bibliography}

\end{document}